\documentclass[12pt,journal,hidelinks,draftclsnofoot,onecolumn]{IEEEtran}\IEEEoverridecommandlockouts
\usepackage{epsfig,graphicx,subfigure,psfrag,amsmath,cases}
\usepackage{latexsym,amssymb,amsmath,epsfig,subfigure,algorithm,mathtools}
\usepackage{algorithmic}
\usepackage{color}
\usepackage{url}
\usepackage{scrtime}
\usepackage{cite}
\usepackage{epstopdf}
\usepackage{hyperref}
\usepackage{booktabs}
\usepackage{multirow}
\usepackage{tabularx}
\newcommand{\tabitem}{~~\llap{\textbullet}~~}

\newtheorem{T-Prob}{Transformed Problem}

\date{\thistime,\,\today}

\begin{document}
\title{A Survey of Downlink Non-orthogonal Multiple Access for 5G Wireless Communication Networks}
\author{\authorblockN{Zhiqiang Wei\IEEEauthorrefmark{1}, Jinhong Yuan\IEEEauthorrefmark{1}, Derrick Wing Kwan Ng\IEEEauthorrefmark{1}, \\Maged Elkashlan\IEEEauthorrefmark{2}, and Zhiguo Ding\IEEEauthorrefmark{3}}

\IEEEauthorrefmark{1} The University of New South Wales, Sydney, Australia\\
\IEEEauthorrefmark{2}Queen Mary University of London, London, UK\\
\IEEEauthorrefmark{3}Lancaster University, Lancaster, UK}
\maketitle

\begin{abstract}
Non-orthogonal multiple access (NOMA) has been recognized as a promising multiple access technique for the next generation cellular communication networks. In this paper, we first discuss a simple NOMA model with two users served by a single-carrier simultaneously to illustrate its basic principles. Then, a more general model with multicarrier serving an arbitrary number of users on each subcarrier is also discussed. An overview of existing works on performance analysis, resource allocation, and multiple-input multiple-output NOMA are summarized and discussed. Furthermore, we discuss the key features of NOMA and its potential research challenges.
\end{abstract}
%

\section{Introduction and Background}
The fifth generation (5G) communication systems is on its way. It is widely believed that 5G is not just an incremental version of the fourth generation (4G) communication systems \cite{Andrews2014} due to not only the increasing demand of data traffic explosion, but also the expected new services and functionalities, such as internet-of-things (IoT), cloud-based architectural applications, etc\cite{Wunder2014a}. These envisioned services pose challenging requirements for 5G wireless communication systems, such as much higher data rates ($100\sim1000 \times$ of current 4G technology), lower latency (1 ms for a roundtrip latency), massive connectivity and support of diverse quality of service (QoS) ($10^6 \;\mathrm{devices/km}^2$ with diverse QoS requirements) \cite{Andrews2014}. From a technical perspective, to meet the aforementioned challenges, some potential technologies, such as massive multiple-input multiple-output (MIMO) \cite{Marzetta2010,Zhu2016MassiveMIMO}, millimeter wave\cite{Pi2011,Rappaport2013}, ultra densification and offloading \cite{Andrews2012,Andrews2013,Ramasamy2013}, have been discussed extensively. Besides, it is expected to employ a future radio access technology for 5G, which is flexible, reliable\cite{DerrickEERobust}, and efficient in terms of energy and spectrum\cite{DerrickEEOFDMA,DerrickEESWIPT}. Radio access technologies for cellular communications are characterized by multiple access schemes, such as frequency-division multiple access (FDMA) for the first generation (1G), time-division multiple access (TDMA) for the second generation (2G), code-division multiple access (CDMA) used by both 2G and the third generation (3G), and orthogonal frequency-division multiple access (OFDMA) for 4G. All these conventional multiple access schemes are categorized as orthogonal multiple access (OMA) technologies, where different users are allocated to orthogonal resources in either time, frequency, or code domain in order to mitigate multiple access interference (MAI). However, OMA schemes are not sufficient to support the massive connectivity with diverse QoS requirements. In fact, due to the limited degrees of freedom (DoF), some users with better channel quality have a higher priority to be served while other users with poor channel quality have to wait to access, which leads to high unfairness and large latency. Besides, it is inefficient when allocating DoF to users with poor channel quality. In this survey, we focus on one promising technology, non-orthogonal multiple access (NOMA), which in our opinion will contribute to disruptive design changes on radio access and address the aforementioned challenges of 5G.
\begin{figure}[ptb]
\centering
\includegraphics[width=0.8\textwidth]{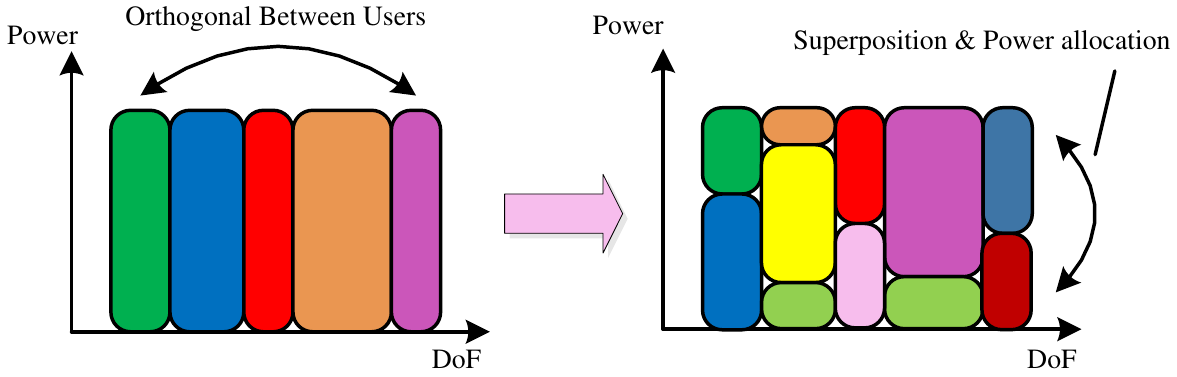}
\caption{From OMA to NOMA via power domain multiplexing.}%
\label{NOMAPrinciple}%
\end{figure}

In contrast to conventional OMA, NOMA transmission techniques intend to share DoF among users via superposition and consequently need to employ multiple user detection (MUD) to separate interfered users sharing the same DoF, as illustrated in Fig. \ref{NOMAPrinciple}. NOMA is beneficial to enlarge the number of connections by introducing controllable symbol collision in the same DoF. Therefore, NOMA can support high overloading transmission and further improve the system capacity given limited resource (spectrum or antennas). In addition, multiple users with different types of traffic request can be multiplexed to transmit concurrently on the same DoF to improve the latency and fairness. The comparison of OMA and NOMA is summarized in Table \ref{NOMAvsOMA}. As a result, NOMA has been recognized as a promising multiple access technique for the 5G wireless networks due to its high spectral efficiency, massive connectivity, low latency, and high user fairness \cite{Dai2015}. For example, multiuser superposition transmission (MUST) has been proposed for the third generation partnership project long-term evolution advanced (3GPP-LTE-A) networks\cite{Access2015}. Three kinds of non-orthogonal transmission schemes have been proposed and studied in the MUST study item. Through the system-level performance evaluation, it has been shown that the MUST can increase system capacity as well as improve user experience.

\begin{figure}[ptb]
\centering
\includegraphics[width=0.8\textwidth]{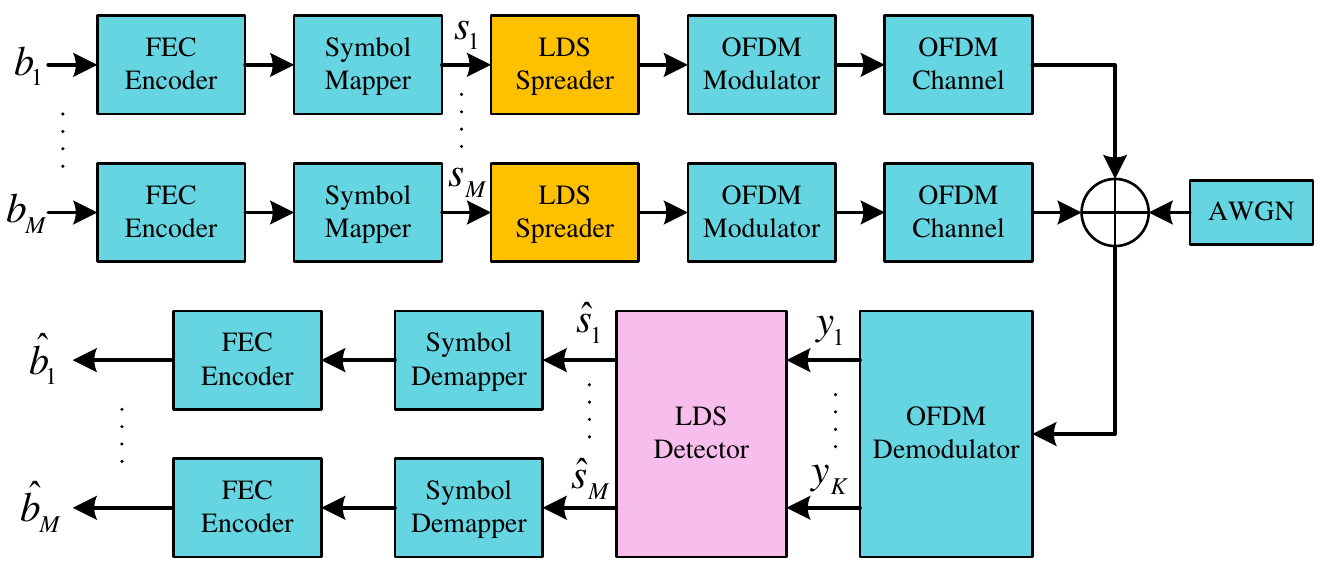}
\caption{Block diagram of an uplink LDS-OFDM system.}%
\label{LDSOFDMBLOCK}%
\end{figure}

Recently, several NOMA schemes have been proposed and received significant attention. According to the domain of multiplexing, the authors in \cite{Dai2015} divided the existing NOMA techniques into two categories, i.e., code domain multiplexing (CDM) and power domain multiplexing (PDM). The CDM-NOMA techniques, including low-density spreading (LDS)\cite{Hoshyar2008,Hoshyar2010,Razavi2012}, sparse code multiple access (SCMA)\cite{Nikopour2013}, pattern division multiple
access (PDMA)\cite{Dai2014PDMA}, etc, introduce redundancy via coding/spreading to facilitate the users separation at the receiver. For instance, LDS-CDMA\cite{Hoshyar2008} intentionally arranges each user to spread its data over a small number of chips and then interleave uniquely, which makes optimal MUD affordable at receiver and exploits the intrinsic interference diversity. LDS-OFDM\cite{Hoshyar2010,Razavi2012}, as shown in Fig. \ref{LDSOFDMBLOCK}, can be interpreted as a system which applies LDS for multiple access and OFDM for multicarrier modulation. Besides, SCMA is a generalization of LDS methods where the modulator and LDS spreader are merged. On the other hand, PDM-NOMA exploits the power domain to serve multiple users in the same DoF, and performs successive interference cancellation (SIC) at users with better channel conditions.

In fact, the non-orthogonal feature can be introduced either in the power domain only or in the hybrid code and power domain. Although CDM-NOMA has the potential code gain to improve spectral efficiency, PDM-NOMA has a simpler implement since there is almost no big change in the physical layer procedures at the transmitter side compared to current 4G technologies. In addition, PDM-NOMA paves the way for flexible resource allocation via relaxing the orthogonality requirement to improve the performance of NOMA, such as spectral efficiency\cite{Hanif2016,sun2016optimal}, energy efficiency\cite{Sun2015a}, and user fairness\cite{Timotheou2015}. Therefore, this paper will focus on the PDM-NOMA, including its basic concepts, key features, existing works, and future research challenges.

\begin{table}[t]
\center
\caption{Comparison of OMA and NOMA.}
  \begin{tabular}{lll}
  \hline
    & Advantages & Disadvantages \\ \hline
  \multirow{3}{*}{OMA}
    & \tabitem Simpler receiver detection & \tabitem Lower spectral efficiency\\
    &  & \tabitem Limited number of users\\
    &  & \tabitem Unfairness for users\\\hline
  \multirow{5}{*}{NOMA}
    & \tabitem Higher spectral efficiency & \tabitem Increased complexity of receivers \\
    & \tabitem Higher connection density & \tabitem Higher sensitivity to channel uncertainty\\
    & \tabitem Enhanced user fairness & \\
    & \tabitem Lower latency & \\
    & \tabitem Supporting diverse QoS & \\\hline
  \end{tabular}
\label{NOMAvsOMA}
\end{table}

\section{Fundamentals of NOMA}
This section presents the basic model and concepts of single-antenna downlink NOMA. The first subsection presents a simple downlink single-carrier NOMA (SC-NOMA) system serving two users simultaneously, while the second subsection presents a more general multi-carrier NOMA (MC-NOMA) model for serving an arbitrary number of users in each subcarrier.

\begin{figure}[ptb]
\centering
\includegraphics[width=0.7\textwidth]{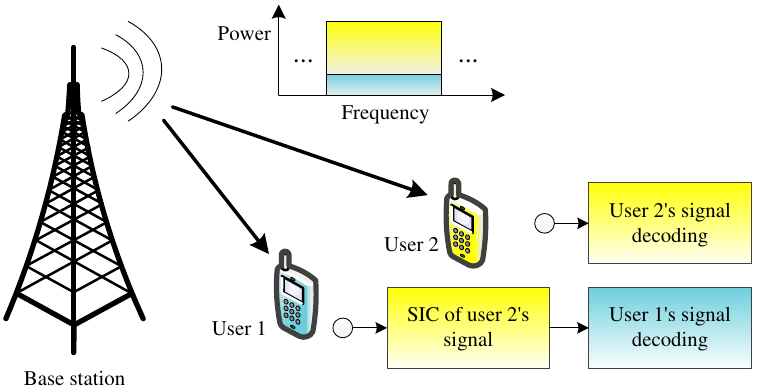}
\caption{A downlink NOMA model with one base station and two users.}%
\label{Downlink_NOMA_model}%
\end{figure}

\subsubsection{Two-user SC-NOMA}\footnote{In this paper, a two-user NOMA systems means that two users are multiplexed on each subcarrier simultaneously. Similarly, a multiuser NOMA systems means that an arbitrary number of users are multiplexed on each subcarrier simultaneously.}
Benjebbour and Saito et al. \cite{Benjebbour2013,Saito2013} proposed the system model of downlink NOMA with superposition transmission at the base station (BS) and successive interference cancellation (SIC) at the user terminals, which can be illustrated in Fig. \ref{Downlink_NOMA_model} in case of one BS and two users.

The BS transmits the messages of both user 1 and user 2, i.e., $s_1$ and $s_2$, with different transmit powers $p_1$ and $p_2$, on the same subcarrier, respectively. The corresponding transmitted signal is represented by
\begin{equation}\label{TWO_UE_TRANSMISSION}
  x = \sqrt {{p_1}} {s_1} + \sqrt {{p_2}} {s_2},
\end{equation}
where transmit power is constrained by $p_1+p_2 = 1$. The received signal at user $i$ is given by
\begin{equation}\label{TWO_UE_RECEIVE1}
  {y_i} = {h_i}x + {v_i},
\end{equation}
where $h_i$ denotes the complex channel coefficient including the joint effect of large scale fading and small scale fading. Variable ${v_i}$ denotes the additive white Gaussian noise (AWGN), and ${v_i} \sim \mathcal{CN}\left( {0,\;\sigma _i^2} \right)$, where $\mathcal{CN}\left( {0,\;\sigma _i^2} \right)$ denotes the circularly symmetric complex Gaussian distribution with mean zero and variance $\sigma _i^2$. Assume that user 1 is the cell-center user with a better channel quality (\emph{strong user}), while user 2 is the cell-edge user with a worse channel quality (\emph{weak user}), i.e., $\frac{{{{\left| {{h_1}} \right|}^2}}}{{\sigma _1^2}} \ge \frac{{{{\left| {{h_2}} \right|}^2}}}{{\sigma _2^2}}$. According to the NOMA protocol\cite{Ding2014}, the BS will allocate more power to the weak user to provide fairness and facilitate the SIC process, i.e., ${p_1} \le {p_2}$.

\begin{figure}[t]
\centering
\includegraphics[width=0.7\textwidth]{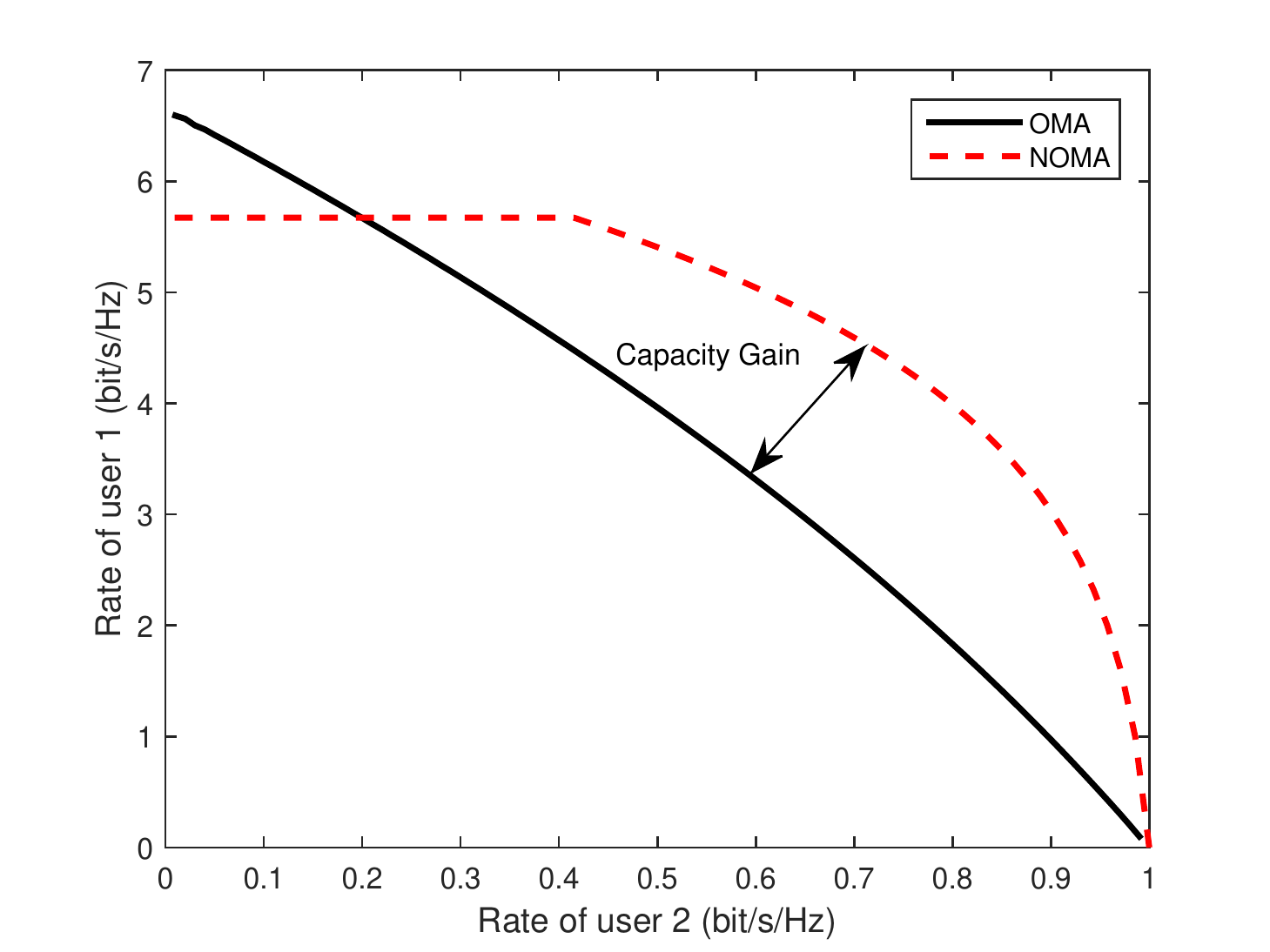}
\caption{The rate region of two-user SC-NOMA in comparison with that of OMA. User 1 is strong user with $\frac{{{{\left| {{h_1}} \right|}^2}}}{{\sigma _1^2}} = 100$, while user 2 is weak user with $\frac{{{{\left| {{h_2}} \right|}^2}}}{{\sigma _2^2}} = 1$.}%
\label{NOMA_over_OMA}%
\end{figure}

In downlink SC-NOMA, the SIC process is implemented at the receiver side. The optimal SIC decoding order is in the descending order of channel gains normalized by noise.
It means that user 1 will decode ${{s}_2}$ first and remove the inter-user interference of user 2 by subtracting ${{s}_2}$ from the received signal ${y_1}$ before decoding its own message ${{s}_1}$. On the other hand, user 2 does not perform interference cancellation and directly decodes its own message ${{s}_2}$ with interference from user 1. Fortunately, the power allocated to user 2 is larger than that of user 1 in the aggregate received signal ${y_2}$, which will not introduce much performance degradation compared to allocating user 2 on this subcarrier exclusively. The rate region of SC-NOMA is illustrated in Fig. \ref{NOMA_over_OMA} in comparison with that of OMA, where it has been proved that NOMA schemes are very likely to outperform OMA schemes in \cite{Xu2015}. Note that the rate region of NOMA only covers a part of the capacity region of broadcast channel with SIC receiver \cite{Tse2005} due to the power constraint ${p_1} \le {p_2}$.

\subsubsection{Multiuser MC-NOMA}
For a downlink MC-NOMA system with one BS serving an arbitrary number of users, such as $K$, the available bandwidth is divided into a set of $N$ subcarriers, where $N<K$, i.e., an overloading scenario that OFDMA cannot afford. The channel between user $k$ and the BS on subcarrier $n$ is denoted by ${h_{k,n}}$, and is assumed to be perfectly known at both transmitter and receiver side. The BS schedules all users across all subcarriers by ${\xi _k}$ and ${\zeta _n}$, where ${\xi _k}$ denotes a user set allocated on subcarrier $k$ and ${\zeta _n}$ denotes a subcarrier set occupied by user $n$. Without loss of generality, the channel gains of all users allocated on subcarrier $k$ are sorted as ${\left| {{h_{k,b\left( 1 \right)}}} \right|^2} \ge {\left| {{h_{k,b\left( 2 \right)}}} \right|^2} \ge  \cdots \ge {\left| {{h_{k,b\left( \left| {{\xi _k}} \right| \right)}}} \right|^2}$, where $\left| {{\xi _k}} \right|$ denotes the cardinality of the user set $\xi _k$ and $b\left(  \cdot  \right)$ indicates the mapping between the sorted channel gain order and the original one. For instance, for subcarrier $k$ occupied by three users ${\xi _k} = \left\{ {1,2,3} \right\}$ and ${\left| {{h_{k,2}}} \right|^2} \ge {\left| {{h_{k,3}}} \right|^2} \ge {\left| {{h_{k,1}}} \right|^2}$, we will have $b\left(  1  \right)=2$, $b\left(  2  \right)=3$, and $b\left(  3  \right)=1$, respectively. Note that the mapping functions are different on different subcarriers due to the different frequency selective fading patterns of different users.

According to NOMA protocol\cite{Ding2014}, all users in ${\xi _k}$ share subcarrier $k$ by different transmission power ${p_{k,b\left( l \right)}}$ based on the given channel gain, where $l = 1,2, \cdots ,\left| {{\xi _k}} \right|$ and ${p_{k,b\left( 1 \right)}} \le {p_{k,b\left( 2 \right)}} \le  \cdots  \le {p_{k,b\left( {\left| {{\xi _k}} \right|} \right)}}$. The sharing strategy saves the subcarriers those might be wasted by only transmitting the messages of the weak users and accommodates more users with diverse QoS requirements, which is favorable to massive connectivity and IoT in 5G networks.

All messages of users in ${\xi _k}$ are superimposed on subcarrier $k$, where the transmitted signal is given by
\begin{equation}\label{Subband_message_superposition}
  {x_k} = \sum\limits_{l = 1}^{\left| {{\xi _k}} \right|} {\sqrt {{p_{k,b\left( l \right)}}} } {s_{k,b\left( l \right)}},
\end{equation}
where ${s_{k,b\left( l \right)}}$ and ${p_{k,b\left( l \right)}}$ denote the message and allocated power of user $n\left(  l  \right)$ on subcarrier $k$, respectively.

Assuming the independent and identically distributed (IID) AWGN over all subcarriers and all users for simplicity, the user scheduling, power allocation, and the SIC decoding order only depends on the channel gain order. At the receiver side, the received signal of user ${b\left( l \right)}$ on subcarrier $k$ can be represented by
\begin{align}\label{Subband_message_received}
{y_{k,b\left( l \right)}} &= {h_{k,b\left( l \right)}}{x_k} + v\\
&= {h_{k,b\left( l \right)}}\sum\limits_{l' = 1}^{\left| {{\xi _k}} \right|} {\sqrt {{p_{k,b\left( {l'} \right)}}} } {s_{k,b\left( {l'} \right)}} + v,\;\forall l \in \left\{ {1,2, \cdots ,\left| {{\xi _k}} \right|} \right\},
\end{align}
where ${v}$ denotes the AWGN, i.e., ${v} \sim CN\left( {0,\;\sigma ^2} \right)$, and $\sigma ^2$ denotes the noise power.

On subcarrier $k$, the scheduled users in ${\xi _k}$ will perform SIC to eliminate inter-user interference. Similar to the case of two-user NOMA, the optimal SIC decoding order is in the descending channel gain order, i.e., $\left\{ {b\left( 1 \right),b\left( 2 \right), \cdots ,b\left( {\left| {{\xi _k}} \right|} \right)} \right\}$. It means that the user ${b\left( l \right)}$ first decodes and subtracts the message ${s_{k,b\left( {l'} \right)}},\;\forall l' > l$, in descending order from ${\left| {{\xi _k}} \right|}$ to $l+1$, and then decodes its own message ${s_{k,n\left( {l} \right)}}$ by treating ${s_{k,n\left( {l'} \right)}},\;\forall l' < l$, as interference.

\section{Performance and Key Features of NOMA}
In this section, we first present the performance characteristics of NOMA in existing works, and then discuss the pros and cons of NOMA schemes.
\subsection{Performance of NOMA}
It has been shown that NOMA offers considerable performance gain over OMA in terms of spectral efficiency and outage probability\cite{Otao2012,Saito2013,Saito2013a,Xu2015,Ding2014,Dingtobepublished}. Initially, the performance of NOMA was evaluated through simulations given perfect CSI by utilizing the proportional fairness scheduler\cite{Otao2012,Saito2013}, fractional transmission power allocation (FTPA)\cite{Saito2013}, and tree-search based transmission power allocation (TTPA)\cite{Saito2013a}. These works showed that the overall cell throughput, cell-edge user throughput, and the degrees of proportional fairness achieved by NOMA are all superior to those of OMA. In \cite{Xu2015}, the author analyzed a two-user SC-NOMA system under statistical CSI from an information theoretic perspective, where it proved that NOMA outperforms native TDMA with high probability in terms of both the sum rate and individual rates.
In \cite{Ding2014}, for a fixed power allocation, the performance of a multiuser SC-NOMA system in terms of outage probability and ergodic sum rates under statistical CSI was investigated in a cellular downlink scenario with randomly deployed users. With the proposed asymptotic analysis, it showed that user $n$ experiences a diversity gain of $n$ and NOMA is asymptotically equivalent to the opportunistic multiple access technique.
Furthermore, the authors in \cite{Yang2016} analyzed the performance degradation of a multiuser SC-NOMA system on outage probability and average sum rates due to partial CSI. It showed that NOMA based on second order statistical CSI always achieves a better performance than that of NOMA based on imperfect CSI, while it can achieve similar performance to the NOMA with perfect CSI in the low SNR region.

In summary, most of the existing works on performance analysis of NOMA focused on a SC-NOMA system since the user scheduling in MC-NOMA complicates the analysis due to its combinatorial nature. A remarkable work in \cite{Dingtobepublished} characterized the impact of user pairing on the performance of a two-user SC-NOMA system with fixed power allocation and cognitive radio inspired power allocation, respectively. The authors proved that, for fixed power allocation, the performance gain of NOMA over OMA increases when the difference in channel gains between the paired users becomes larger. However, further exploration on performance analysis of MC-NOMA system should be carried out in the future since user scheduling is critical for performance of NOMA.

\subsection{Pros}
\begin{enumerate}
  \item Higher spectral efficiency: By exploiting the power domain for user multiplexing, NOMA systems are able to accommodate more users to cope with system overload. In contrast to allocate a subcarrier exclusively to a single user in OMA scheme, NOMA can utilize the spectrum more efficiently by admitting strong users into the subcarriers occupied by weak users without compromising much their performance via utilizing appropriate power allocation and SIC techniques.
  \item Better utilization of heterogeneity of channel conditions: As we mentioned before, NOMA schemes intentionally multiplex strong users with weak users to exploit the heterogeneity of channel condition. Therefore, the performance gain of NOMA over OMA is larger when channel gains of the multiplexed users become more distinctive \cite{Dingtobepublished}.
  \item Enhanced user fairness: By relaxing the orthogonal constraint of OMA, NOMA enables a more flexible management of radio resources and offers an efficient way to enhance user fairness via appropriate resource allocation\cite{Timotheou2015}.
  \item Applicability to diverse QoS requirements: NOMA is able to accommodate more users with different types of QoS requests on the same subcarrier. Therefore, NOMA is a good candidate to support IoT which connects a great number of devices and sensors requiring distinctive targeted rates.
\end{enumerate}

\subsection{Cons}
\begin{enumerate}
  \item The BS needs to know the perfect channel state information (CSI) to arrange the SIC decoding order, which increases the CSI feedback overhead.
  \item The SIC process introduces a higher computational complexity and delay at the receiver side, especially for multicarrier and multiuser systems.
  \item The strong users have to know the power allocation of the weaker users in order to perform SIC, which also increases the system signalling overhead.
  \item Allocating more power to the weak users, who are generally in the cell-edge, will introduce more inter-cell interferences into the whole system.
\end{enumerate}

\section{Design of NOMA Schemes}
Due to the remarkable performance gain of NOMA over conventional OMA, a lot of works on design of NOMA schemes have been proposed in literatures. In this section, we present the existing works on resource allocation of NOMA and MIMO-NOMA, and then briefly introduce other works associated with NOMA.
\subsection{Resource Allocation}
Resource allocation has received significant attention since it is critical to improve the performance of NOMA. However, optimal resource allocation is very challenging for MC-NOMA systems, since user scheduling and power allocation couple with each other severely. Some initial works on resource allocation in \cite{Otao2012,Saito2013,Saito2013a} have been reported, but they are far from optimal. In \cite{Hojeij2015,Hojeij2015a}, the authors studied a two-user MC-NOMA system by minimizing the number of subcarriers assigned under the constraints of maximum allowed transmit power and requested data rates, and further introduced a hybrid orthogonal-nonorthogonal scheme.
Furthermore, the authors in \cite{sun2016optimal} studied a joint power and subcarrier allocation problem for a two-user MC-NOMA system. They proposed an optimal scheme and a suboptimal scheme with close-to-optimal performance based on monotonic optimization and difference of convex function programming, respectively.

Besides, there are also several works on resource allocation for multiuser MC-NOMA systems. In \cite{Di2015}, the authors formulated the resource allocation problem to maximize the sum rate, which is a non-convex optimization problem due to the binary constraint and the existence of the interference term in the objective function. Interestingly, they proposed a suboptimal solution by employing matching theory and water-filling power allocation.
In \cite{Lei2015}, the authors presented a systematic approach for NOMA resource allocation from a mathematical optimization point of view. They formulated the joint power and channel allocation problem of a downlink multiuser MC-NOMA system, and proved its NP-hardness based on \cite{Liu2014} via defining a special user. Furthermore, they proposed a competitive suboptimal algorithm based on Lagrangian duality and dynamic programming, which significantly outperforms OFDMA as well as NOMA with FTPA.

Most of works aforementioned focus on the optimal resource allocation for maximizing the sum rate. However, fairness is another objective to optimize for resource allocation of NOMA. Proportional fairness (PF) has been adopted as a metric to balance the transmission efficiency and user fairness in many works\cite{Kim2004PF,Wengerter2005PF}. In \cite{Liu2015b}, the authors proposed a user pairing and power allocation scheme for downlink two-user MC-NOMA based on the PF objective. A prerequisite for user pairing was given and a closed-form optimal solution for power allocation was derived. Apart from PF, max-min or min-max methods are usually adopted to achieve user fairness. Given a preset user group, the authors in \cite{Timotheou2015} studied the power allocation problem from a fairness standpoint by maximizing the minimum achievable user rate with instantaneous CSI and minimizing the maximum outage probability with average CSI. Although the resulting problems are non-convex, simple low-complexity algorithms were developed to provide close-to-optimal solutions. Similarly, another paper \cite{Shi2015} studied the outage balancing problem of a downlink multiuser MC-NOMA system to maximize the minimum weighted success probability with and without user grouping. Joint power allocation and decoding order selection solutions were given, and the inter-group power and resource allocation solutions were also provided.

In summary, many existing works focus on the resource allocation for NOMA systems under perfect CSI at the transmitter side. However, there are only few works on the joint user scheduling and power allocation problem for MC-NOMA systems under imperfect CSI, not to mention the SIC decoding order selection problem. In fact,
under imperfect CSI, the SIC decoding order cannot be determined by channel gain order, and some other metrics, such as distance, priority, and target rates, are potential criteria to decide the SIC decoding order.

\begin{figure}[ptb]
\centering
\includegraphics[width=0.8\textwidth]{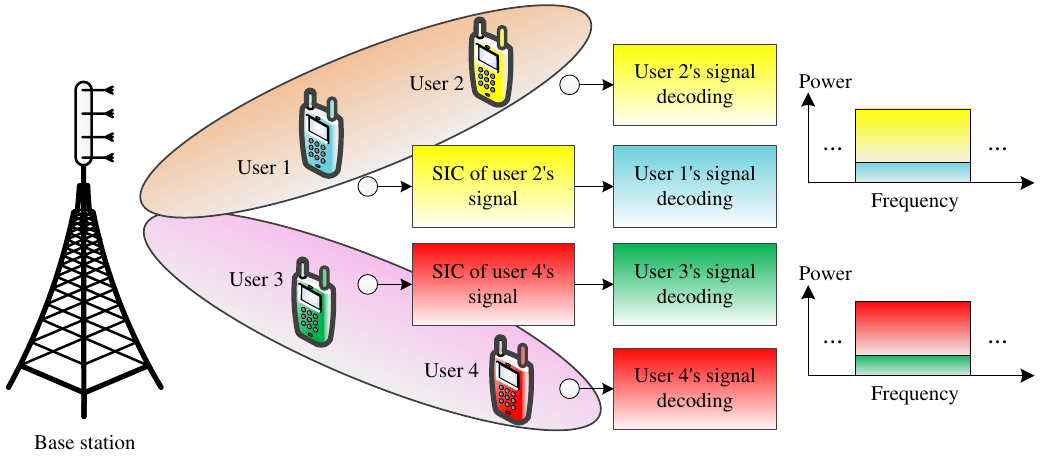}
\caption{A downlink MIMO-NOMA model with one base station and four users.}%
\label{Downlink_MIMO_NOMA_model}%
\end{figure}
\subsection{MIMO-NOMA}
The application of MIMO techniques to NOMA systems is important to enhance the performance gains of NOMA. Therefore, MIMO-NOMA is another hot topic that has been researched, where the BS and users are equipped with multiple antennas, and multiple users in the same beam are multiplexed on power domain. Fig. \ref{Downlink_MIMO_NOMA_model} illustrates a simple MIMO-NOMA system with one base station and four users. Initially, the concept of MIMO-NOMA was proposed in \cite{Saito2013a,Lan2014,Chen2014}, which demonstrated that MIMO-NOMA outperforms conventional MIMO OMA. The authors in \cite{Kim2013} proposed a two-user MIMO-NOMA scheme with a clustering and power allocation algorithm, where the correlation and channel gain difference were taken into consideration to reduce intra-beam interference and inter-beam interference simultaneously. In \cite{Choi2015}, the authors proposed a minimum power multicast beamforming scheme and applied to two-user NOMA systems for multi-resolution broadcasting. The proposed two-stage beamforming method outperforms the zero-forcing beamforming scheme in \cite{Kim2013}.

The design of precoding and detection algorithms also received considerable attention since they are the key to eliminate or reduce the inter-cluster interferences.
The authors in \cite{Sun2015} studied the ergodic sum capacity maximization problem of a two-user MIMO-NOMA system under statistical CSI with the total power constraint and minimum rate constraint for the weak user. This paper derived the optimal input covariance matrix, and proposed the optimal power allocation scheme as well as a low complexity suboptimal solution. Furthermore, in \cite{Sun2015b}, the authors studied the sum rate optimization problem of two-user MIMO-NOMA under perfect CSI with the same constraints, while different precoders were assigned to different users. The optimal precode covariance matrix was derived by utilizing the duality between uplink and downlink, and a low complexity suboptimal solution based on singular value decomposition (SVD) was also provided. In \cite{Dingtobepublisheda}, the authors proposed a new design of precoding and detection matrices for a downlink multiuser MIMO-NOMA system, then analyzed the impact of user pairing as well as power allocation on the sum rate and outage probability of MIMO-NOMA system. Furthermore, in \cite{Ding2015a}, a transmission framework based on signal alignment was proposed for downlink and uplink two-user MIMO-NOMA systems. The authors in \cite{Hanif2016} studied the sum rate maximization problem of a downlink multiuser multiple-input single-output (MISO) NOMA system. They showed that MISO NOMA transmission outperforms conventional OMA schemes, particularly when the transmit SNR is low, and the number of users is greater than the number of BS antennas. Recently, a multiuser MIMO-NOMA scheme based on limited feedback was proposed and analyzed in \cite{Ding2015c}.

In summary, most of the existing works on MIMO-NOMA focused on design of precoding and detection algorithms, and their performance analyses. However, user scheduling and power allocation were rarely discussed in the spatial domain, which play important roles to improve the spatial efficiency of MIMO-NOMA.

\subsection{Other Works on NOMA}
In addition to the above two aspects, there are many other works associated NOMA. We will not discuss further in detail due to the limited space. Compared to downlink NOMA, uplink NOMA was also studied in several works\cite{Takeda2011,Al-Imari2014,Chen2015,Chen2015a,Al-Imari2015,Kim2015,Zhangtobepublished}. Moreover, asynchronous NOMA has also been investigated in uplink scenarios\cite{Haci2015,Haci2015a}. Cooperative NOMA, where strong users serve as relays for weak users, was studied in \cite{Ding2015,Ding2016a}. In addition, several works on NOMA combined with other techniques were also reported, such as energy harvesting \cite{Liu2015a,Diamantoulakis2015}, cognitive radio networks \cite{Liu2016}, visible light communication\cite{Marshoud2015}, and physical layer security\cite{Zhang2016SSR}.

\section{Research Challenges}
As discussed above, NOMA can be employed to improve the spectral efficiency, user fairness, as well as to support massive connections with diverse QoS requirements. Based on our overview of existing works on NOMA and its potential applications in practical systems, we present the research challenges of NOMA in the following three aspects.
\subsection{Resource Allocation under Imperfect CSI}
Most of existing works on resource allocation of NOMA are based on the assumption of perfect CSI at the transmitter side, which is difficult to obtain in practice due to either the estimation error or the feedback delay. Therefore, it is nature to investigate how CSI error affects the performance of NOMA and to consider robust resource allocation under imperfect CSI\cite{Wei2016NOMA}. Since NOMA is expected to offer lower latency in order to support delay-sensitive applications in 5G, one promising solution is the outage-based robust approach for designing the resource allocation of NOMA.
In this direction, the SIC decoding order under imperfect CSI is still an open problem. Furthermore, it is important to study the joint optimization of power allocation, user scheduling and SIC decoding order selection of NOMA under imperfect CSI.

\subsection{Cooperative NOMA}
A key feature of NOMA is that the strong users have prior information of the weak users, which has not been fully exploited in existing works. In cooperative NOMA, the strong users can serve as relays for the weak users, which has the potential to utilize the spatial DoF even for users with a single antenna. Some preliminary works showed that cooperative NOMA can achieve the maximum diversity gain for all the users \cite{Ding2015,Ding2016a}. It is important to study the optimal resource allocation for cooperative NOMA. Besides, distributed beamforming can be employed in cooperative NOMA to harvest the spatial DoF without much signalling overhead. Considering that cooperative NOMA will introduce more complexity and extra delay into systems, it is important to investigate the tradeoffs among the system performance, complexity, and delay.
\subsection{QoS-based NOMA}
As we mentioned before, NOMA has great potential to support diverse QoS requirements. The heterogeneity of QoS requirements might in turn facilitate the power allocation and user scheduling of NOMA, which is also an interesting topic to explore in the future. For example, users in NOMA systems can be categorized according to their QoS requirements, instead of their channel conditions, which offers two following benefits. One is that the SIC decoding order, power allocation, and user scheduling can be designed more appropriately to meet the users¡¯ QoS requests. The other is to make NOMA communications more general, e.g., applicable to scenarios in which users¡¯ channel conditions are the same.

\section{Conclusion}
In this article, a promising multiple access technology for 5G networks, NOMA, is discussed. A two-user SC-NOMA scheme and a multiuser MC-NOMA scheme were presented and discussed to illustrate the basic concepts and principles of NOMA.
A literature review about performance analyses of NOMA, resource allocation for NOMA, and MIMO-NOMA were discussed. Furthermore, we presented the key features and potential research challenges of NOMA.

\bibliographystyle{IEEEtran}
\bibliography{LiteratureReview}

\end{document}